\documentclass[a4paper]{jpconf}
\def \BE {\begin{equation}}
\def \EE {\end{equation}}
\def \BEA { \begin{eqnarray}}
\def \EEA {\end{eqnarray}}
\def \BEAH { \begin{eqnarray*}}
\def \EEAH {\end{eqnarray*}}
\def \bl {\mbox{\boldmath{$\ell$}}}
\def \bk {\mbox{\boldmath{$k$}}}
\def \hbl {\mbox{\boldmath{$\hat \ell$}}}
\def \bn {\mbox{\boldmath{$n$}}}
\def \hbn {\mbox{\boldmath{$\hat n$}}}
\def \hbm #1 {\mbox{\boldmath{$\hat m^{(#1)}$}}}
\def \btm {\mbox{\boldmath{$m$}}}
\def \cbm {\mbox{\boldmath{$\bar m$}}}
\def \bm #1 {\mbox{\boldmath{$m^{(#1)}$}}}
\def \hbm #1 {\mbox{\boldmath{$\hat m^{(#1)}$}}}
\def \BDM {\begin{displaymath}}
\def \EDM {\end{displaymath}}
\newcommand{\lp}{\left(}
\newcommand{\rp}{\right)}

\begin{document}
\title{On the algebraic classification of spacetimes}
\author{V. Pravda}
\address{Mathematical Institute, Academy of Sciences, \v{Z}itn\'{a} 25, 115 67 Prague
1, Czech Republic}
\ead{pravda@math.cas.cz}
\begin{abstract}
We briefly overview the Petrov classification in four dimensions and its generalization to higher dimensions. 
\end{abstract}

\section{Introduction - the Petrov classification in four dimensions}

The Petrov classification, which was developed by Petrov \cite{Petrov} and extended  e.g. in \cite{Geh}-\cite{Penrose60}
(see \cite{Pen-Rind}-\cite{Hall} for additional references), invariantly determines algebraic type of the Weyl tensor at a given point
of a spacetime. 
We say that a spacetime is of a given Petrov type if the type is same at all points. There are several equivalent methods of  obtaining
the classification. The most elegant one is probably the two component spinor approach developed by Penrose {\cite{Penrose60,Pen-Rind}}.
In this approach the Weyl tensor is represented by a totally symmetric spinor $\Psi_{ABCD}$ and consequently it can
be decomposed in terms of four principal spinors $\alpha_{A}$, $\beta_{B}$, $\gamma_C$, $\delta{_D}$
$$
\Psi_{ABCD} = \alpha_{(A} \beta_{B} \gamma_C \delta_{D)} . 
$$
Petrov types now correspond to various multiplicities of principal spinors and following possibilities occur
\BEAH
I=\{1111\}:&&\ \  \Psi_{ABCD} = \alpha_{(A} \beta_{B} \gamma_C \delta_{D)} ,  \\
II=\{211\}:&&\ \  \Psi_{ABCD} = \alpha_{(A} \alpha_{B} \gamma_C \delta_{D)} ,  \\
D=\{22\}:&&\ \  \Psi_{ABCD} = \alpha_{(A} \alpha_{B} \beta_C \beta_{D)}  , \\
III=\{31\}:&&\ \  \Psi_{ABCD} = \alpha_{(A} \alpha_{B} \alpha_C \beta_{D)} ,  \\
N=\{4\}:&&\ \  \Psi_{ABCD} = \alpha_{A} \alpha_{B} \alpha_C \alpha_{D} .  \\
O=\{-\}:&&\ \  \Psi_{ABCD} = 0
\EEAH
All Petrov types except of the type I are called algebraically special.

Null vectors corresponding to  various principal spinors via $v^{\alpha}= \xi^A {\bar \xi}^{\dot A}$ are  called {\it principal null vectors}
and corresponding directions {\it principal null directions} (PNDs).
Thus there are at most four distinct PNDs at each point of a spacetime. 
In the tensorial form, PND corresponding to a vector $\bl$  satisfies three 
equivalent conditions \cite{Pen-Rind,Kramer} given in
Table 1 for various Petrov types.
\begin{table}
\begin{tabular}{rll}
$\ell^b \ell^c \ell_{[e}C_{a]bc[d}\ell_{f]}=0$ &   $\ell$  at least simple PND (I) & $\Psi_0=0$, $\Psi_1 \not=0$\\
$\ell^b \ell^c C_{abc[d}\ell_{e]}=0$&  $\ell$  at least double PND (II,D) & $\Psi_0=0$, $\Psi_1=0$, $\Psi_2 \not=0$\\ 
$ \ell^c C_{abc[d}\ell_{e]}=0$ &   $\ell$  at least triple PND (III) & $\Psi_0=0$, $\Psi_1=0$, $\Psi_2=0$, $\Psi_3 \not=0$\\ 
$\ell^c C_{abcd}=0$ &  $\ell$   quadruple PND (N) & $\Psi_0=0$, $\Psi_1=0$, $\Psi_2=0$, $\Psi_3=0$, $\Psi_4 \not=0$\\
\label{PND4D}
\end{tabular}
\caption{Each line contains three equivalent conditions for various Petrov types in four dimensions. $\Psi_0 \dots \Psi_4$ are complex
components of the Weyl tensor in the Newman-Penrose formalism \cite{NP,Pen-Rind,Kramer}. The third condition means that it is {\it possible}
to choose a frame in which given components of the Weyl tensor vanish.}
\end{table}

In some cases there is a
close correspondence between the Petrov type and symmetries or a physical interpretation of the spacetime. For example
a static spherically symmetric spacetime is of type D \cite{Wald}, with 
two double PNDs defining radially ingoing and outgoing null congruences.    Similarly one would expect 
that  Friedmann-Robertson-Walker spacetime does not contain any preferred null direction and consequently it is of type O \cite{Pen-Rind}
and that the simplest gravitational waves (such as pp-waves) have only one preferred null direction and thus are of type N. 
(In fact  gravitational waves with a longitudinal component which decays faster than transverse type N waves are of type III.)

In order to find exact solutions of the Einstein equations it is usually necessary to assume some symmetries. Another possibility
is to assume a specific algebraically special type of a spacetime since then the field equations can also be considerably simplified.

In some cases the Petrov type can be distinct at some points or regions from the rest of a spacetime. An interesting example
are black holes of type I  with isolated or Killing horizons of type II \cite{Ashtekar,BH4D}.

As it was mentioned  above, in four dimensions there are several distinct methods leading to the Petrov classification and only some of them can be generalized 
to higher
dimensions. Let us  briefly describe  the  appropriate method, which can be also used in higher dimensions, at first in four dimensions: 

In four dimensions in the standard complex null tetrad (${\bl}$, $\bn$, $\btm$,  $\cbm$) \cite{Pen-Rind,Kramer} the Weyl tensor has five complex
 components
$\Psi_0 \dots \Psi_4$. A boost in $\bl - \bn$ plane is described by  ${\hbl}=\lambda \bl$, ${\hbn}=\lambda^{-1} \bn$ and
we say that a quantity $q$ has a {\it boost weight} $b$ if it transforms under boost according to $\hat q = \lambda^b q$.
Components of the Weyl tensor $\Psi_0$, $\Psi_1$, $\Psi_2$, $\Psi_3$, $\Psi_4$ have boost weight -2,-1,0,1,2, respectively.
When $\bl$ coincides with PND then $\Psi_0$ vanishes. Other components can be transformed away only in algebraically special cases.
Namely, as can be seen from the Table 1, for types I, II, III, N one can transform away components with boost weight greater or
 equal to 2,1,0,-1,  respectively. Type D is a special subcase of type II, all components with non-zero boost weight can be transformed
 away. In the next section we will use a similar approach for the classification of the Weyl tensor in higher dimensions.

\section{Classification of the Weyl tensor in higher dimensions}

In this section we overview the classification of the Weyl tensor in higher ($D$) dimensions \cite{Algclass,Weylletter}. 
In higher dimensions it is more convenient to use a real frame 
\BDM
\bm{0} 
= \bn, \ \bm{1} 
=\bl, \ \bm{i} 
 , \ \ \ i,j,k = 2 \dots D-1,
\EDM
with two null vectors   $\bn,\ \bl$  
\BDM
\ell^a \ell_a= n^a n_a = 0,\ \   \ell^a n_a = 1, \ \  a = 0  \dots D-1,
\EDM
 and $D-2$ spacelike vectors $\bm{i} $ 
\BEAH
 m^{(i)a}m^{(j)}_a=\delta_{ij},\ \ m^{(i)a}\ell_a=0=m^{(i)a}n_a,  \ \ \ \ \ \ i,j,k = 2 \dots D-1. 
\EEAH

The metric has the form
\BDM
g_{a b} = 2 \ell_{(a}n_{b)} + \delta_{ij} m^{(i)}_a m^{(j)}_b ,  \label{metric} 
\EDM
which remains unchanged under null rotations
\BDM
\hbl =  \bl +z_i {\bm{i} } -\frac{1}{2} z^i z_i\, \bn , \ \    
\hbn =  \bn, \ \ 
    \hbm{i} =  \bm{i} - z_i \bn ,
    \label{nullrot}
\EDM
 { spins}  
\BDM
\hbl =  \bl, \ \ \hbn = \bn, \ \hbm{i} =  X^{i}_{\ j} \bm{j} ,\ \ \ \ \ \ \ \ \ \ 
\EDM
and {boosts}
\BDM
\hbl = \lambda \bl, \ \  \hbn = \lambda^{-1} \bn, \ \ \hbm{i} = \bm{i} .
\EDM

Let us define the {\it boost order} of a tensor ${\mathbf T}$   as the maximum boost weight of
its frame components. It can be shown that the { boost order} of a tensor  depends only on the choice of a
null direction  $\bl$ and thus spins and boost do not affect it \cite{Algclass}. 
We will denote the boost order of a given tensor ${\mathbf T}$ as $b(\bl)$ indicating the dependence
on the choice of $\bl$. We suppose that the particular choice of ${\mathbf T}$ is clear from
the context (in our case it is the Weyl tensor).

Let us denote the maximum value 
of   $b({\bk})$   taken over all null vectors ${\bk}$ as  $b_{\rm{max}}$. 
Then we say that a null vector   ${\bk}$   is {\it aligned}  with the tensor   ${\mathbf T}$  
whenever    $b({\bk}) < b_{\rm{max}}$. The integer   $b_{\rm{max}}-b({\mathbf k})-1$
  is called {\it order of alignment}. 

For the Weyl tensor aligned null vectors (or directions) represent a natural generalization
  of the principal null directions and we will call them WANDs (Weyl aligned null directions).

 The classification of the Weyl tensor in higher dimensions can be based  on the existence of WANDs
  of various orders of alignment. The  Weyl tensor in an arbitrary dimension has in general components
with boost weights $-2 \leq b \leq2$  and thus
 the order of alignment of a WAND cannot exceed 3.

We say that the {\it primary alignment type} of the Weyl tensor is G if there are no WANDs
 and it is 1, 2, 3, 4 if the maximally aligned null
vector has order of alignment 0, 1, 2, 3, respectively.

Once we fix $\bl$ as a WAND
with  maximal order of alignment, we can search for $\bn$ with maximal order
of alignment subject to the constraint $\bn \cdot \bl=1$   and similarly define {\it secondary
alignment type}. Alignment type is a pair consisting of primary and secondary alignment types.
Possible alignment types are summarized in Table 2. We also introduce  {\it Weyl type} with
notation emphasizing the link with the four dimensional Petrov classification.

\begin{table}[h]
\begin{center}
\begin{tabular}{|cc|c|}
\hline
\ \ D$>$4 dimensions & & 4 dimensions \\
\hline
Weyl type & alignment type& Petrov type   \\
\hline
G     & G &      \\
I     & (1)   &  \\
${\rm I}_{i}$ & (1,1) & I \\
II    & (2)  &  \\
${\rm II}_{i}$ & (2,1) & II \\
D & (2,2) & D \\
III & (3) &  \\
${\rm III}_{i}$ & (3,1) & III \\
N & (4) & N \\
\hline
\end{tabular}
\caption{Classification of the Weyl tensor in four and higher dimensions. Note that in four dimensions
alignment type (1) is necessarily equivalent to the type (1,1), (2) to (2,1) and (3) to (3,1) and since there is always at least one  PND, 
type G does not exist.} 
\end{center}
\end{table}

In order to express the Weyl tensor in terms of its components in the $\bl$, $\bn$, $\bm{i} $ frame we introduce an operation \{ \ \}
\BDM
w_{\{a} x_b y_c z_{d\}}  \equiv \frac{1}{2} (w_{[a} x_{b]} y_{[c} z_{d]}+ w_{[c} x_{d]} y_{[a} z_{b]}).
\label{zavorka}
\EDM
Now  the Weyl tensor in arbitrary dimension can be written as
\BEAH
  C_{abcd} &=& 
  \overbrace{
    4 C_{0i0j}\, n^{}_{\{a} m^{(i)}_{\, b}  n^{}_{c}  m^{(j)}_{\, d\: \}}}^2  
  +\overbrace{
    8C_{010i}\, n^{}_{\{a} \ell^{}_b n^{}_c m^{(i)}_{\, d\: \}} +
    4C_{0ijk}\, n^{}_{\{a} m^{(i)}_{\, b} m^{(j)}_{\, c} m^{(k)}_{\, d\: \}}}^1  
  \nonumber \\&& {
    \begin{array}{l} +
      4 C_{0101}\, \, n^{}_{\{a} \ell^{}_{ b} n^{}_{ c} \ell^{}_{\, d\: \}} 
\;  + \;  C_{01ij}\, \, n^{}_{\{a} \ell^{}_{ b} m^{(i)}_{\, c} m^{(j)}_{\, d\: \}}  \\[2mm]
      +8 C_{0i1j}\, \, n^{}_{\{a} m^{(i)}_{\, b} \ell^{}_{c} m^{(j)}_{\, d\: \}}
   +  C_{ijkl}\, \, m^{(i)}_{\{a} m^{(j)}_{\, b} m^{(k)}_{\, c} m^{(l)}_{\, d\: \}}
    \end{array} \Biggr\}^0
    }
  \label{eq:rscalars}\\ 
   &+& \overbrace{
    8 C_{101i}\, \ell^{}_{\{a} n^{}_b \ell^{}_c m^{(i)}_{\, d\: \}} +
    4 C_{1ijk}\, \ell^{}_{\{a} m^{(i)}_{\, b} m^{(j)}_{\, c} m^{(k)}_{\, d\: \}}}^{-1} 
  + \overbrace{
      4 C_{1i1j}\, \ell^{}_{\{a} m^{(i)}_{\, b}  \ell^{}_{c}  m^{(j)}_{\, d\: \}}}^{-2} .
\EEAH
Note that summation over $i$, $j$, $k$, $l$ indices is implicitly assumed and that boost orders of various components are also indicated. 

Number of independent frame components of various boost weights,
is
\BDM
\fl
\overbrace{2 \lp \frac{(m+2)(m-1)}{2} \rp}^{2,-2} + \overbrace{2 \lp \frac{(m+1)m(m-1)}{3} \rp}^{1,-1} 
+\overbrace{\frac{m^2(m^2-1)}{12} + \frac{m(m-1)}{2}}^0 ,
\EDM
with $m=D-2$. This is in agreement with the number of independent components of the Weyl tensor
\BDM
  \label{eq:wdim}
  \frac{(D+2)(D+1)D(D-3)}{12}  .
\EDM

In the general case, the Weyl tensor is  quite complicated, however, in some algebraically special cases
it is much simpler and this can substantially simplify the field equations. The simplest case are type N
spacetimes (such as pp-waves) with only $D (D-3) /2$ independent components and the Weyl tensor of the form
\BDM
C_{abcd} = 4 C_{1i1j}\, \ell^{}_{\{a} m^{(i)}_{\, b}  \ell^{}_{c}  m^{(j)}_{\, d\: \}}.
\EDM
In fact type N in higher dimensions was already identified before the whole classification was worked out \cite{pphd}.
Similarly in \cite{Horowitz} it was proven that static spherically symmetric spacetimes are ``boost invariant''
(this corresponds to the type D) in arbitrary dimension.

In \cite{Lewandowski} it was proven that  in arbitrary dimension 
isolated horizons are locally of type II (or of more special type) similarly as in four dimensions.

The generalization of the Kerr metric for higher dimensions, the Myers-Perry solution, is also of 
type D in five dimensions \cite{Bianchi} (see \cite{CPclass} for other type D black hole solutions in higher dimensions)
 while the black ring is of the type ${\rm I}_{i}$ (and of type II on the horizon) \cite{WandsBR}. 
 
Interestingly, the peeling theorem for weakly asymptotically  simple spacetimes in even dimensions can be, under certain
assumptions, also demonstrated \cite{peeling}.
 
 Note that in higher dimensions  conditions given in the first column of Table 1 are 
only necessary (but not sufficient) conditions for the statements in the second
column \cite{WandsBR}. For example, a spacetime satisfying   $\ell^c C_{abcd}=0$
can be of the type II \cite{WandsBR}. Such a  spacetime has in principle a non-vanishing
curvature invariant $C_{abcd} C^{abcd}$ and for type N and III spacetimes in arbitrary dimension all polynomial
curvature invariants constructed from components of the Weyl tensor vanish.

Let us at the end discuss interesting but not fully explored connection between algebraic type of the Weyl
tensor and consequences following from the Bianchi and Ricci identities.
In accordance with \cite{Bianchi} we decompose first covariant derivatives of $\bl$ as
\BDM
\ell_{a ; b} = L_{11} \ell_a \ell_b + L_{10} \ell_a n_b + L_{1i} \ell_a m^{(i)}_{\, b}  +
L_{i1} m^{(i)}_a \ell_b   + L_{i0} m^{(i)}_{\, a} n_{b} + L_{ij} m^{(i)}_{\, a} m^{(j)}_{\, b}   \label{dl} \\
\EDM
and denote symmetric and antisymmetric parts of $L_{ij}$ as $S_{ij}$ and $A_{ij}$, respectively.
Now we can define the expansion $\theta$ and the shear  $\sigma_{ij}$ as follows:
\BEA 
\theta &\equiv &\frac{1}{n-2} \ell^{a}_{\ ; a} = \frac{1}{n-2}S_{kk} , \nonumber \\
\sigma_{ij} &\equiv& \left( 
\ell_{(a ; b)} 
- \theta \delta_{kl}   m^{(k)}_a m^{(l)}_b \right)
m^{(i)a} m^{(j)b}  
=S_{ij}-\frac{S_{kk}}{n-2}\delta_{ij}. \nonumber \label{shearmatrix}
\EEA

In vacuum, where the Riemann tensor is equal to the Weyl tensor, the Bianchi identities $R_{abcd;e}+R_{abde;c}+R_{abec;d}=0$
for type N spacetimes imply 
\BEAH
L_{k[i} \Psi_{j]k}=0, \\
 \Psi_{i[j}L_{k]0}=0,  \\
  L_{k[j} \Psi_{m]i} + L_{i[m} \Psi_{j]k}=0,  \\
 \Psi_{ik}A_{jm} + \Psi_{ij}A_{mk} + \Psi_{im}A_{kj} =0,
 \EEAH 
where $\Psi_{ij}=C_{1i1j}/2$. From these equations it follows that in arbitrary dimension the congruence corresponding to the WAND 
is geodesic \cite{Bianchi}. This is also true in four dimensions. However, it also follows that in higher dimensions for vacuum type N 
spacetimes with  non-vanishing expansion the matrices $S_{ij}$, $A_{ij}$ and $\Psi_{ij}$ have certain specific form \cite{Bianchi} and that 
shear does not vanish. According to the four-dimensional Goldberg-Sachs theorem \cite{Goldberg, Kramer} the multiple principal null vector in 
vacuum algebraically special spacetimes is geodesic and shearfree. From the above mentioned results we can see
that the Goldberg-Sachs theorem does not have a straightforward generalization for higher dimensions (this was also pointed out in \cite{Frolov}). 
However, some partial steps towards the generalization can be done (see Sec. 5 in \cite{Bianchi}).

Let us conclude that the algebraic classification of  higher dimensional spacetimes gives us a new important tool in higher 
dimensional gravity. As the standard four-dimensional Petrov classification, it allows us to classify and study physical properties of spacetimes 
and hopefully it will also generate new algebraically special solutions.

\end{document}